\documentclass{article}
\usepackage{graphicx}
\usepackage{amsmath,amssymb,amsfonts,latexsym}

\newtheorem{theorem}{Theorem}[section]

\newtheorem{proposition}[theorem]{Proposition}

\newtheorem{problem}[theorem]{Problem}

\newcommand{\nvec}{{\bf n}}
\newcommand{\imag}{\, \text{Im}\,}
\newcommand{\pvec}{{\bf p}}
\newcommand{\qvec}{{\bf q}}
\newcommand{\rvec}{{\bf r}}
\newcommand{\svec}{{\bf s}}
\newcommand{\uvec}{{\bf u}}
\newcommand{\vvec}{{\bf v}}
\newcommand{\wvec}{{\bf w}}
\newcommand\sgn{\text{sgn}\,}
\newcommand{\Rcal}{{\cal R}}
\newcommand{\Rr}{{\mathbb R}}

\newcommand{\Zz}{{\mathbb Z}}

\newcommand{\Pcal}{{\cal P}}
\newcommand{\Ncal}{{\cal N}}
\newcommand\wbar{{\bar w}}
\newcommand\tbar{{\bar t}}
\newcommand{\evec}{{\bf e}}
\newcommand{\half}{{\textstyle \frac{1}{2}}}
\newcommand\grad{{\bf \nabla}}
\newcommand\inff{\text{inf}}
\newcommand\avec{{\bf a}}
\newcommand\bvec{{\bf b}}
\newcommand{\nuvec}{{\boldsymbol \nu}}
\newcommand\Omegaabs{W}
\newcommand\Phat{{\hat P}}
\newcommand\xhat{{\hat x}}
\newcommand\yhat{{\hat y}}
\newcommand\zhat{{\hat z}}

\newcommand\adj{\,\text{adj}\,}
\begin{document}

\title{Liquid crystals and harmonic maps in polyhedral domains}
\author{Apala Majumdar, Jonathan Robbins and Maxim Zyskin} 

\maketitle

\begin{abstract}
  Unit-vector fields $\nvec$ on a convex polyhedron $P$ subject to
  tangent boundary conditions provide a simple model of nematic
  liquid crystals in prototype bistable displays.  The
  equilibrium and metastable configurations correspond to  minimisers and local
  minimisers of the Dirichlet energy, and may be regarded as $S^2$-valued
  harmonic maps on $P$.  We consider unit-vector fields which
  are continuous away from the vertices of $P$.
  A lower bound for the infimum Dirichlet energy for a given homotopy
  class is obtained as a sum of minimal connections between fractional
  defects at the vertices of $P$.  In certain cases, this lower bound
  can be improved by incorporating certain nonabelian homotopy
  invariants.  For a rectangular prism, upper bounds for the infimum
  Dirichlet energy are obtained from locally conformal solutions of
  the Euler-Lagrange equations, with the ratio of the upper and lower
  bounds bounded independently of homotopy type.  However, since the
  homotopy classes are not weakly closed, the infimum may not be
  realised; the existence and regularity properties of continuous
  local minimisers of given homotopy type are open questions.
  Numerical results suggest that some homotopy classes always contain
  smooth minimisers, while others may or may not depending on the
  geometry of $P$.  Numerical results modelling a bistable device
  suggest that the observed nematic configurations may be
  distinguished topologically.\\
  
\noindent  {\it This article appears as  a chapter in ``Analysis and
  Stochastics of Growth Processes and Interface Models'', P Morters et
  al.~eds., Oxford University Press 2008,
  http://www.oup.com/uk/catalogue/?ci=9780199239252.}

\end{abstract}


\section{Introduction}\label{sec:introduction}
Liquid crystals are intermediate phases of matter exhibiting partial
ordering in the orientation and/or positions of their constituent
particles.  The constituents of nematic liquid crystals have a
distinguished axis, and in the nematic phase these axes tend to align.
The direction and degree of alignment can exhibit a rich variety of
singularities.  Standard references on liquid crystals include 
\cite{degennes, virga, klemanlav, stewart}.

\begin{figure}
\begin{center}
\centering
\includegraphics*[totalheight=0.2\textheight,
viewport = 280 153 420 290,clip
]{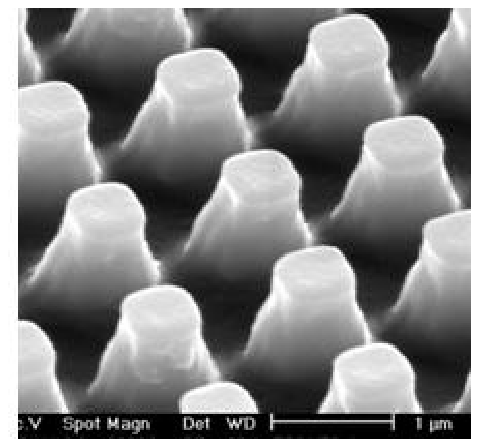}
\caption{The PABN cell 
(from \cite{kg2002})}
\label{fig: pabn_photo}
\end{center}
\end{figure}

The nematic phase is optically birefringent (light propagation is
polarisation-dependent).  This, together with the fact that nematic
ordering can be modified by external electric and magnetic fields, has
led to a wide range of display applications.  Most
present-day liquid crystal displays (eg twisted nematic) are based on
monostable cells, where, in the absence of external fields, the
orientation assumes a single (spatially varying) equilibrium
configuration which is effectively transparent to incident polarised
light.  To produce and maintain optical contrast, voltage pulses,
which change the orientation, must be continually applied.  There is
considerable interest in developing  bistable cells, which
support two (and possibly more) stable  configurations
with contrasting optical properties.  In bistable cells, power is
needed only to switch between  configurations.  One mechanism
for engendering bistability is to introduce microstructures into the
geometry 
\cite{jones,kg2002, mottram2007}.
Nematic
liquid crystals in cells with polyhedral features (eg, ridges, posts,
wells) have been found to support multiple configurations.  One such device,
the PABN, or post-aligned bistable nematic cell, is shown in Fig.\ 
\ref{fig: pabn_photo} 
(\cite{kg2002}).
It consists of a liquid
crystal layer sandwiched between two planar substrates, with the lower
substrate featured by an array of microscopic posts.

As a simple model for such systems, we consider nematic liquid
crystals in a convex polyhedron $P \subset \Rr^3$ with orientation
described by a director field, $\nvec:P \rightarrow RP^2$, taking
values in the real projective plane.  We
consider the case of strong azimuthal anchoring,  described by
{\it tangent boundary conditions}. Tangent boundary conditions require
that, on a face of $P$, $\nvec$ lies tangent to the face, but is
otherwise unconstrained.  It follows  that on the edges of $P$,
$\nvec$ is parallel to the edges, and therefore is necessarily
discontinuous at the vertices.  We are interested as to whether
equilibria can be classified according to homotopy, and therefore 
restrict our attention to director fields which are continuous away
from the vertices.  For these, we can unambiguously assign an
orientation to the director field (as $P$ is simply connected), and
regard $\nvec$ as a unit-vector field.  We let $C_T^0(P,S^2)$ denote
the space of continuous unit-vector fields on $P$ satisfying tangent
boundary conditions, or tangent unit-vector fields for short.

The elastic or Oseen-Frank energy of a configuration $\nvec$ is given
by
\begin{multline}
  \label{eq:elastic}
  E = \int_P \big [ K_1 (\grad\cdot \nvec)^2 + K_2 (\nvec
  \cdot (\grad \times \nvec))^2 +   K_3
  (\nvec\times (\grad \times \nvec))^2\\ +  K_4 \grad\cdot ((\nvec\cdot
  \grad)\nvec - (\grad\cdot \nvec) \nvec)\big ]\, dV.
\end{multline}
Tangent boundary conditions 
imply that the contribution from the $K_4$-term,
which is a pure divergence, vanishes. We shall make use of the so-called one-constant approximation, in which the
remaining elastic constants $K_1$, $K_2$ and $K_3$ are taken to be the
same and set to unity.  In this case, (\ref{eq:elastic}) becomes the
Dirichlet energy,
\begin{equation}
  \label{eq:one-constant}
  E(\nvec) =  \int_P  (\grad \nvec)^2\, dV.
\end{equation}
Minimisers of the Dirichlet energy, which correspond to equilibrium
configurations, are $S^2$-valued harmonic maps, as are
local minimisers, which correspond to metastable configurations. 

The homotopy classes of $C_T^0(P,S^2)$ are described in
Section~\ref{sec:homot-class}, and a lower bound for the infimum of
the Dirichlet energy in each homotopy class is given in
Section~\ref{sec:lower-bound-dirichl}.  The lower bound is expressed
as a sum of minimal connections between fractional defects at the
vertices of $P$, in analogy with the well-known result of
\cite{brecorlie} 
for the infimum Dirichlet energy of a set of point
defects in $\Rr^3$.  For nonconformal homotopy classes, this bound can
be improved by incorporating certain nonabelian homotopy invariants;
this is shown explicitly for certain homotopy classes in a rectangular
prism in Section\ \ref{sec:nonab}.  Unlike the case of point defects
in $\Rr^3$, the lower bound of Section~\ref{sec:lower-bound-dirichl}
is expected to be strictly less than the infimum; achieving the lower
bound would require concentration along a minimal connection, which
would be incompatible with tangent boundary conditions.  However, for
$P$ a rectangular prism, we can construct trial configurations in each
homotopy class whose energies differ from the lower bound by a factor
which is bounded independently of $h$
(Section~\ref{sec:upper-bound-prism}).  Generalising the construction
to arbitrary $P$ requires finding conformal maps on $S^2$ which
preserve a given set of geodesics.

It is an open question as to whether the infimum is
achieved in a given homotopy class, as is the regularity of the local
minimisers.  Numerical results presented in
Section~\ref{sec:exist-regul-minim} suggest that some homotopy classes
always contain smooth minimisers, while others may or may not
depending on the geometry of $P$.  Numerical results for a model of a
bistable display suggests that the observed nematic configurations are
topologically distinct.

In addition to existence and regularity questions, it would be
interesting to investigate   namics under the influence of applied
fields.  Switching between configurations of different homotopy type
requires the creation and destruction of defects, and one would like
to understand this process in detail.

\section{Homotopy classification}\label{sec:homot-class} 

\begin{figure}
\begin{center}
\centering
\includegraphics[totalheight=0.26\textheight,
viewport= 
20 20 575 437, clip
]{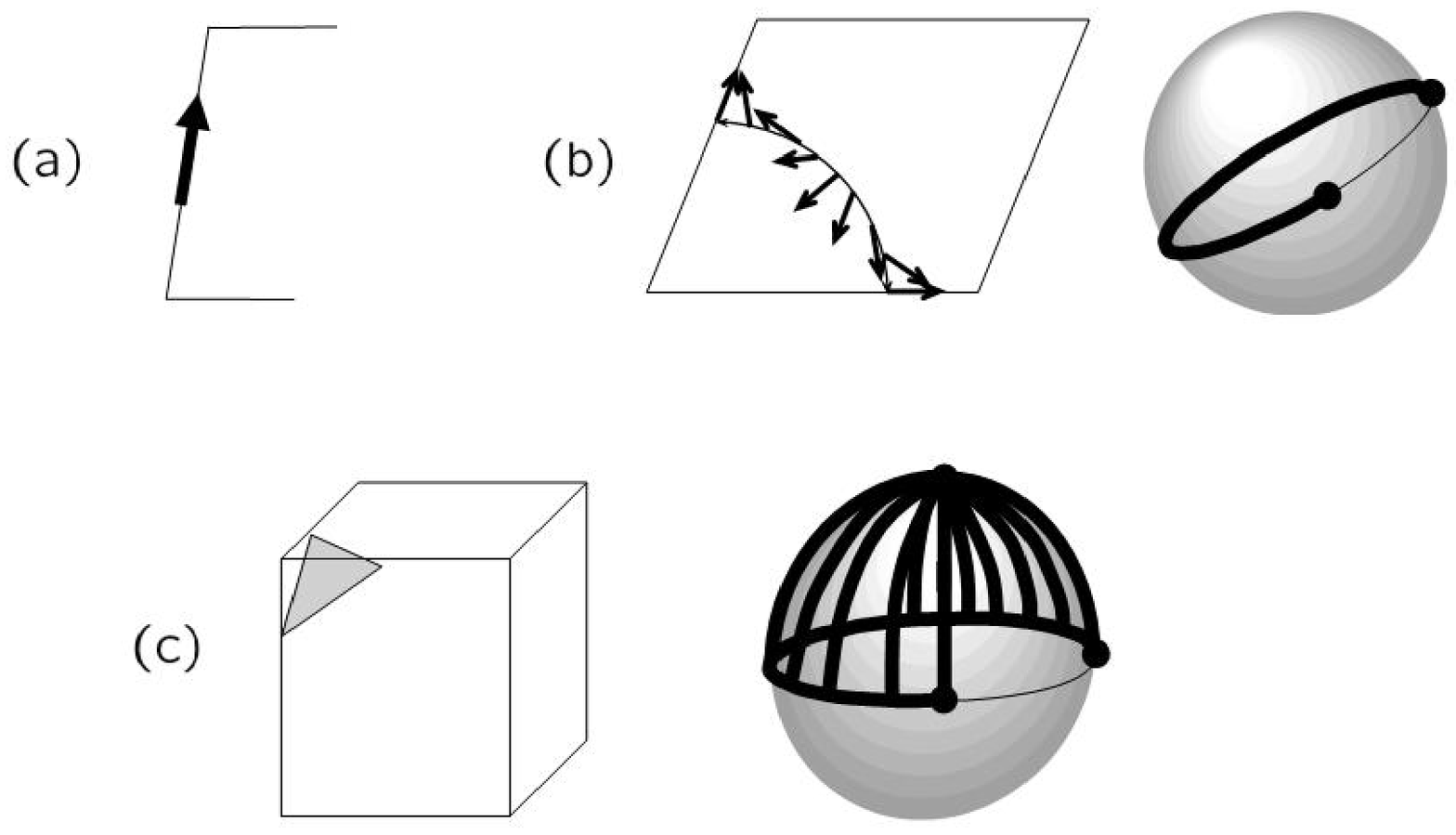}
\caption{Homotopy invariants. (a) Edge orientation (b) Kink number.
  $\nvec$ describes a 3/4-turn about the vertex (the image on
  $S^2$ is also shown), corresponding to kink number $-1$. (c) Trapped
  area.  The image of the cleaved surface on $S^2$ has signed area $-3\pi/2$.}
\label{fig: invariants}
\end{center}
\end{figure}

Given $\nvec \in C_T^0(P,S^2)$ we can identify a number of
discrete-valued quantities which depend continuously on $\nvec$ and
which are therefore homotopy invariants. (Details may be found in
\cite{rz2003} and \cite{mrz2007}).  
Along an edge of $P$,
$\nvec$ must lie parallel to the edge, so its value there is
determined up to a sign, which we call an {\it edge orientation} (see
Fig.\ \ref{fig: invariants}(a)).  Next, along a path on a face of $P$
between two edges, $\nvec$ must lie tangent to the face, and therefore
describes a geodesic on $S^2$, ie an arc of a great circle
(see Fig.\ \ref{fig: invariants}(b)).
As the endpoints of the geodesic are fixed by the edge orientations,
the geodesic may be assigned an integer-valued relative winding
number, or {\it kink number}.  By convention, the shortest geodesic is
assigned kink number zero.
Another invariant is associated with a surface which separates one of
the vertices of $P$ from the other vertices -- we call this a {\it cleaved
  surface}  (see Fig.\ \ref{fig: invariants}(c)).  Along the boundary of a cleaved surface, $\nvec$ is
determined up to homotopy by its edge orientations and kink numbers.
Therefore, the signed area of the image of the cleaved surface itself,
called the {\it trapped area} at the vertex, is determined up to an
integer multiple of $4\pi$ (ie, some number of whole coverings of the
sphere).

Collectively, the edge orientations, kink numbers and trapped areas
constitute a complete set of homotopy invariants for $C_T^0(P,S^2)$;
two configurations are homotopic if and only if their invariants are
the same.  We note that the invariants are not all
independent -- continuity of configurations 
on the faces of $P$ implies that the kink numbers on
each face satisfy a sum rule, while continuity 
on the interior of $P$ implies that the trapped areas add
up to zero.  One can show that every set of invariants satisfying
these sum rules can be realised.

\begin{figure}
\begin{center}
\centering
\includegraphics[totalheight=0.2\textheight,
viewport= 75 167 469 357,clip
]{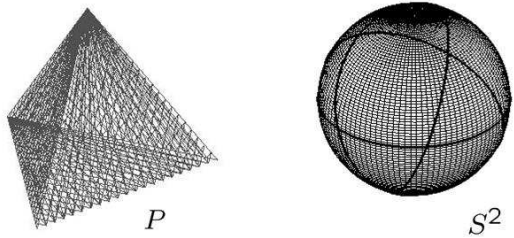}
\caption{Sectors for a tetrahedron.  The  great circles of
  directions tangent to the four faces partition the two-sphere into
  14 sectors.}
\label{fig: sectors}
\end{center}
\end{figure}

From the preceding discussion, it is evident that the invariants of
$\nvec$ can be determined from its values on a set of cleaved surfaces
(the values of $\nvec$ on the corners and edges of the cleaved
surfaces determine its edge orientations and kink numbers).  Given a set of
cleaved surfaces we can define an alternative set of invariants, the
{\it wrapping numbers}, which will be used in subsequent sections. 
 Let
$T \subset S^2$ denote the set of directions which are tangent to
one of the faces of $P$.
Then $T$ consists of a union of great circles.
$S^2 - T$ consists of a union of disjoint open spherical polygons,
which we call {\it sectors} (see Fig.\ \ref{fig: sectors}).
Let $C^a$ denote a cleaved surface separating the
$a$th vertex of $P$, say, from the others, and let $\nvec^a$ denote the restriction of
$\nvec$ to $C^a$. 
Let $\Sigma^\sigma$ denote the $\sigma$th sector of
$S^2$. 
The wrapping number $w^{a\sigma}$ is the number of times
$\nvec^a$ covers $\Sigma^\sigma$, counted with orientation.  For $\nvec$
differentiable, this is given by
\begin{equation}
  \label{eq:wrapping_def_as_integal}
  w^{a\sigma} 
 = \frac{1} {A^\sigma} \int_{C^a} \,
\nvec^*(\chi^\sigma \omega), 
\end{equation}
where $\omega$ is the area two-form on
$S^2$, normalised to have integral $4\pi$, $\chi^\sigma$ is the
characteristic function of $\Sigma^\sigma$, $\nvec^*$ denotes the
pull-back, and
$  A^\sigma = \left| 
\int_{S^2} \chi^\sigma \,\omega 
\right |$
is the   area
of $\Sigma^\sigma$.
Alternatively, $w^{a\sigma}$ can be expressed as
the index of a
regular value $\svec \in \Sigma^\sigma$, ie
\begin{equation}
  \label{eq:wrapping numbers}
  w^{a\sigma} = \sum_{\rvec\, | \, \nvec^a(\rvec) = \svec} \sgn \det
  (\nvec^a)^{\bf '}(\rvec).
\end{equation}
The wrapping numbers are homotopy invariants, and using Stokes'
theorem can be expressed in
terms of the edge orientations, kink numbers and trapped areas.
These relations can also be inverted to obtain 
the edge orientations, kink numbers and trapped areas in terms of the
wrapping numbers.  Thus, the
wrapping numbers constitute a complete (though redundant) set of
homotopy invariants. 

If the nonzero wrapping numbers at a given vertex
are all negative, the homotopy class is said to be {\it conformal}
with respect to that vertex, and if positive, 
{\it
  anticonformal} with respect to that vertex. A homotopy class is
called {\it nonconformal} if there are vertices with  wrapping
numbers of different signs.  

It is straightforward to count the number of invariants
as well as the relations among them. 
Suppose that $P$ has $f$ faces, $e$ edges and
$v$ vertices (so that, from Euler's formula, $f  - v + e = 2$).  Then
$P$ has $v$ trapped areas, which satisfy a single sum rule; $2e$ kink
numbers, which satisfy $f$ sum rules; and $e$ edge orientations.  It also
follows from Euler's formula applied to the set of tangent directions $T$, regarded as a graph
on $S^2$, that there are generically (and at most) $f^2-f+2$ sectors
(`generically' means that no direction is tangent to three or more
faces of $P$).  Thus, there are generically (and at most) $(f^2 - f +
2)v$ wrapping numbers, many more than the number of trapped areas and
kink numbers.  Among the constraints on the wrapping numbers, we point
out that for a fixed sector $\Sigma^\sigma$, their sum over vertices must vanish, ie
\begin{equation}
  \label{eq:w_sum_rule}
  \sum_a w^{a\sigma} = 0.
\end{equation}
We will use  $h$ to denote both an admissible set of values of
the invariants
as well as the homotopy class in $C^0_T(P,S^2)$ characterised by 
these values. 

\section{Lower bound: minimal
  connection}\label{sec:lower-bound-dirichl} 
\cite{brecorlie}
established the infimum Dirichlet energy for unit-vector fields on
$\Rr^3$ with point defects of specified position and degree.  The result
is expressed in terms of a {\it minimal connection} between the
defects, defined below.  A similar argument yields a lower
bound for the infimum Dirichlet energy for tangent unit-vector fields
on $P$ of fixed homotopy type, in which the vertices of $P$ play the
role of defects, and the wrapping numbers that of generalised
degrees.  Details may be found in
\cite{mrz2004a}, \cite{mrz2004b}, and \cite{mrz2007}.

We first review the result of 
\cite{brecorlie}.  
Let $\Omega = \Rr^3 -
\{\rvec^1,\ldots,\rvec^n\}$,  and let $\nvec:\Omega\rightarrow S^2$
denote a unit-vector field on $\Omega$.  
Continuous unit-vector fields on
$\Omega$ may be classified up to homotopy by their degrees, $d =
(d^1,\ldots,d^n)\in \Zz^n$, on spheres about each of the excluded
points $\rvec_j$ (the restriction of $\nvec$ to such a sphere may be regarded as
a map from $S^2$ into itself).  For $\nvec$ smooth,
\begin{equation}
  \label{eq:degree}
  d^j = \frac{1}{4\pi} \int_{|\rvec - \rvec^j| = \epsilon}  \nvec^* \omega
\end{equation}
for small enough $\epsilon$.
For $\grad \nvec$
square-integrable, the Dirichlet energy is given as in
(\ref{eq:one-constant}) by
\begin{equation}\label{eq:Omega energy}
   E(\nvec) = \int_\Omega(\grad \nvec)^2 dV.
\end{equation}
In order for $E(\nvec)$ to be finite, we require that
\begin{equation}
  \label{eq:d_sum}
  \sum_j d^j = 0.
\end{equation}
Let $C^0_\Omega(d)$ denote the homotopy class of continuous
unit-vector fields with degrees $d$ satisfying (\ref{eq:d_sum}), and
let 
\begin{equation}
  \label{eq:E_h_BCL}
  E^\inff_\Omega(d) = \inf_{\nvec \in C^0_\Omega(d)\cap H^1(\Omega, S^2)} E(\nvec)
\end{equation}
denote the infimum energy in $C^0_\Omega(d)$.

Given two $m$-tuples of points in $\Rr^3$, $\Pcal = (\avec^1,\ldots, \avec^m)$ and
$\Ncal = (\bvec^1,\ldots,\bvec^m)$ (whose points need not be distinct), a
{\it connection} is a pairing $(\avec^j,\bvec^{\pi(j)})$ of points in $\Pcal$ and $\Ncal$,
specified here in terms of a permutation $\pi \in S_m$ ($S_m$ denotes the
symmetric group).  The length of
a connection is the sum of the distances between the paired points,
and a {\it minimal connection} is a connection of minimum length.  Let
\begin{equation}
  \label{eq:minimal_connection}
  L(\Pcal,\Ncal) = \min_{\pi \in S_m} \sum_{j  = 1}^m |\avec^j - \bvec^{\pi(j)}|
\end{equation}
denote the length of a minimal connection,
and let $|d| = \half
  \sum_j |d^j|$.
\begin{theorem}\label{thm:0}
{\rm \cite{brecorlie}}
  The infimum $E^\inff_\Omega(d)$ of the Dirichlet energy of
  continuous unit-vector fields on the domain $\Omega = \Rr^3 -
  \{\rvec^1,\ldots,\rvec^n\}$ of  degrees $d^j$ about the
  excluded points $\rvec^j$ is given by
  \begin{equation}
    \label{eq:BCL-result}
    E^\inff_\Omega(d) = 8\pi L(\Pcal(d),\Ncal(d)),
  \end{equation}
  where $\Pcal(d)$ is the $|d|$-tuple of excluded points of positive degree, 
with $\rvec^j$ included $d^j$ times, and 
$N(d)$ is the $|d|$-tuple of excluded points of negative degree,
with $\rvec^k$ included $|d^k|$ times.
\end{theorem}
In fact, the result of 
\cite{brecorlie} 
applies to more general
domains with holes.  

Here we sketch the argument that $8\pi
L(\Pcal(d),\Ncal(d))$ 
is a lower bound for
$E^\inff_\Omega(d)$. 
 It suffices to consider
smooth unit-vector fields on $\Omega$, as these are dense in
$C^0_\Omega(d)\cap H^1(\Omega, S^2)$.  For any orthonormal frame
$\uvec$, $\vvec$, $\wvec$, one has the inequality
\begin{equation}
  \label{eq:local_ineq}
  (\grad \nvec)^2 \ge  2 | (d\xi \wedge
  \nvec^*\omega)(\uvec,\vvec,\wvec)| \ge 2 (d\xi \wedge \nvec^*\omega)(\uvec,\vvec,\wvec),
\end{equation}
where $\xi$ is differentiable and $  | d \xi | \le 1$.
(\ref{eq:local_ineq}) follows from the fact that, at every point, there is at least one
direction (say $\uvec$) in which the directional derivative $\nabla_u
\nvec := (\uvec \cdot \grad) \nvec$ 
vanishes, while
\begin{equation}
  \label{eq:ineq_2}
(\nabla_v \nvec)^2 + (\nabla_w \nvec)^2 \ge 2 |\nabla_v \nvec| \,
  |\nabla_w \nvec| \ge 2 | \nabla_v \nvec \wedge \nabla_w \nvec|.
\end{equation}
Since $d\omega = 0$, it follows that $d\xi \wedge \nvec^*\omega =
d(\xi \nvec^* \omega)$, so that 
\begin{equation}
  \label{eq:local_ineq_3}
  (\grad \nvec)^2 \ge 2 d(\xi \nvec^* \omega).
\end{equation}
Substituting (\ref{eq:local_ineq_3}) into (\ref{eq:Omega energy}) 
and applying Stokes' theorem, we get a lower bound
\begin{equation}
  \label{eq:intermed_1a}
  E^\inff_\Omega(d) \ge 2\sum_{j} \xi^j d^j,
\end{equation}
which depends only on the values $\xi^j := \xi(\rvec^j)$ of $\xi$
at the defects. Since $|d\xi| \le 1$, these values are constrained by
$ |\xi^j - \xi^k| \le |\rvec^j - \rvec^k|$.
In fact, every set of $\xi^j$'s satisfying these constraints
can be realised by a piecewise-differentiable
function $\xi$ (eg, let $\xi(\rvec) = \max_j (\xi^j - |\rvec -
\rvec^j|)$).  Thus, one obtains a bound
\begin{equation}
  \label{eq:intermed_1b}
  E^\inff_\Omega(d) \ge 2\max_{\xi^j} \sum_{j} \xi^j d^j, \
  \text{where}\  \xi^j \ge 0, \ |\xi^j - \xi^k| \le |\rvec^j - \rvec^k|,
\end{equation}
in the form of a finite-dimensional linear optimisation problem.

The dual formulation is given by
\begin{equation}
  \label{eq:intermed_1c}
  E^\inff_\Omega(d) \ge 2\min_{\eta_{jk}} \sum_{jk} \eta_{jk}
  |\rvec^j - \rvec^k|,
 \ \text{where}\ \eta_{jk} \ge 0, \sum_k (\eta_{jk} -
\eta_{kj}) \ge d^j.
\end{equation}
This is a sort of transport problem, in which the degrees are the
quantities to be transported and the costs are the distances between
defects.  We can take $\eta_{jk}$ to be $0$ unless $d^j > 0$ and
$d^k < 0$.  Without loss of generality, we may also assume that the
degrees are either $+1$ or $-1$, so that there are an equal number, $m
:= n/2$, of each, with positions $(\avec^1,\ldots, \avec^m)$ and
$(\bvec^1,\ldots,\bvec^m)$ respectively (if not, repeat each defect
according to its multiplicity).  (\ref{eq:intermed_1c}) becomes
\begin{equation}
  \label{eq:intermed_2}
  E^\inff_\Omega(d) \ge 2\min_{M_{pq}} \sum_{p,q = 1}^m M_{pq}
  |\avec^p - \bvec^q|,  \ \text{where}\ M_{pq} \ge 0, \sum_p M_{pq} = \sum_q M_{qp} = 1.
\end{equation}
As $M$ is constrained to be doubly stochastic, a theorem of Birkhoff
\cite{birkhoff} 
implies that it lies in the convex hull of the set of
$m$-dimensional permutation matrices.  The optimal solution will
be amongst the permutation matrices themselves, leading to $
E^\inff_\Omega(d) \ge 8\pi L(\Pcal(d),\Ncal(d))$.

A similar argument leads to a lower bound for the infimum Dirichlet
energy for tangent unit-vector fields on $P$ of given homotopy type.
\begin{theorem}\label{thm:1}
{\rm \cite{mrz2007}}
Let $h = \{w^{a\sigma}\}$ be an admissible topology for continuous tangent
unit-vector fields on a polyhedron $P$. 
The infimum $E^\inff(h)$ of the Dirichlet energy of continuous tangent
unit-vector fields on $P$ with invariants $h$ is bounded below by
\begin{equation}
  \label{eq:E_-}
  E^\inff_P(h) \ge  
\sum_\sigma 2 A^\sigma L(\Pcal^\sigma(h), \Ncal^\sigma(h)),
\end{equation}
where $\Pcal^\sigma$
(resp.~$\Ncal^\sigma$) contains the vertices of $P$  for which
$w^{a\sigma}$ is positive (resp.~negative), each such vertex included
with multiplicity 
$|w^{a\sigma}|$.
\end{theorem}
\noindent Thus, to each sector $\sigma$ may be associated a constellation of point
defects at the vertices $\vvec^a$ with degrees $w^{a\sigma}$.  The
lower bound of (\ref{eq:E_-}) is a sum of the lengths of minimal
connections for these constellations, weighted by the sector areas $A^\sigma$.

\section{Lower bound: nonabelian invariants}\label{sec:nonab}
For nonconformal homotopy classes, the lower bound of
Theorem~\ref{thm:1} can be improved by incorporating certain
nonabelian invariants.  These invariants, and the sense in which they
are nonabelian, are introduced in Section~\ref{sec:absol-degr-spell}
in a two-dimensional setting.  For tangent unit-vector fields on $P
\subset \Rr^3$ we describe this phenomenon in a particular case
(Section~\ref{sec:refl-symm-homot}), reflection-symmetric homotopy
classes in a rectangular prism.  Details will be given in
\cite{mrz2007b}.  

\subsection{Absolute degree and spelling length}\label{sec:absol-degr-spell}
Let $\phi:D^2 \rightarrow \Rr^2$ be a smooth map of the two-disk into
the plane.  We recall that $x \in D^2$ is a regular point of $\phi$ if
$x$ is in the interior of $D^2$ and $\det \phi'(x) \ne 0$, $y \in
\Rr^2$ is regular value of $\phi$ if all of its preimages are regular
points, and a regular value has a finite number of preimages.  Let
$\Rcal(\phi)\subset \Rr^2$ denote the set of regular values of $\phi$.
From Sard's theorem, $\imag \phi - \Rcal(\phi)$ is of zero measure.

Given $y \in \Rcal(\phi)$, the algebraic degree of $y$ (or degree, for short)
is given by
\begin{equation}
  \label{eq:deg_y}
  d_\phi(y) = \sum_{x \in \phi^{-1}(y)} \sgn \det \phi'(x),
\end{equation}
and is 
invariant under smooth deformations of $\phi$
which preserve the boundary map $\partial\phi$.  We define the {\it absolute
  degree} of $y$  by 
\begin{equation}
  \label{eq:Deg_y}
  D_\phi(y) =  \sum_{x \in \phi^{-1}(y)} 1.
\end{equation}
Clearly $D_\phi(y)$  is not invariant under all deformations which
preserve $\partial \phi$, and 
\begin{equation}
  \label{eq:deg_and_Deg}
  D_\phi(y) \ge |d_\phi(y)|.
\end{equation}

Let $R = \{y_1,\ldots,y_n\} \subset \Rcal(\phi)$ denote a set of $n$
regular values of $\phi$.  We may regard the boundary $\partial \phi$
as a map $\partial \phi: S^1 \rightarrow \Rr^2- R$ from the circle to
the $n$-times-punctured plane.  Let $\pi_1(\Rr^2-R,p)$ denote the
fundamental group of $\Rr^2 - R$, based at a point $p$, and let
$[\partial \phi]\in \pi_1(\Rr^2-R,p)$ denote the homotopy class of
$\partial \phi$.

The fundamental group $\pi_1(\Rr^2-R,p)$ may be identified with the
free group on $n$ generators, $F(c_1,\ldots,c_n)$ (see, eg,
\cite{magnus}).  
Let us specify that the generator $c_j$ corresponds to a
loop which encircles $y_j$ once with positive orientation but contains
no other $y_k$'s.  This determines the $c_j$'s up to conjugacy.
Given $g \in \pi_1(\Rr^2-R,p)$, we define a {\it spelling} to be a
factorisation of $g$ into a product of conjugated generators and
inverse generators, eg
\begin{equation}
  \label{eq:spelling}
  g = h_1 c_{i_1}^{\epsilon_1} h_1^{-1} \cdots h_r c_{i_r}^{\epsilon_r} h_r^{-1},
\end{equation}
where $h_s \in  \pi_1(\Rr^2-R,p)$ and $\epsilon_s = \pm 1$.  The
length of a spelling is the number of factors (ie, $r$ in (\ref{eq:spelling})).
Define the {\it spelling length}, denoted 
$\Lambda_n(g)$, to be the shortest possible length of a spelling of $g$
(eg,  $\Lambda_n(c_1 c_2 c_1^{-1}
c_2^{-1}) = 2$).  The spelling length of the identity, $e$, is
taken to be zero.  
It turns out that the spelling length of $[\partial \phi]$ gives a
lower bound on the sum of the absolute degrees of points in $R$.
\begin{proposition}\label{prop:disk}
Given $\phi: D^2\rightarrow \Rr^2$ smooth, $R = \{y_1,\ldots,y_n\} \subset
\Rcal(\phi)$,
and $\pi_1(\Rr^2-R,p) \backsimeq F_n(c_1,\ldots,c_n)$, with generators
$c_j$ as above.
 Then
\begin{equation}
  \label{eq:disk}
  \sum_{j=1}^n D_\phi(y_j) \ge \Lambda_n([\partial \phi]).
\end{equation}
\end{proposition}

Let $\bar F_n(c_1,\ldots,c_n)$ denote the abelianisation of
$F_n(c_1,\ldots,c_n)$, obtained by taking all of the $c_j$'s to
commute, and given $g \in F_n(c_1,\ldots, c_n)$, let $\bar g$ denote
the corresponding element of $\bar F_n(c_1,\ldots,c_n)$.  Then $\bar
g$ can be written as $c_1^{\delta_1} \cdots c_n^{\delta_n}$
for some integers $\delta_j$.  Let $\bar \Lambda_n(g) = \sum_{j=1}^n
|\delta_j| $ .  Clearly $\Lambda_n(g) \ge \bar\Lambda_n(g)$ (eg, $\bar
\Lambda_n(c_1 c_2 c_1^{-1} c_2^{-1}) = 0)$.  It is readily seen that
\begin{equation}
  \sum_{j=1}^n |d_\phi(y_j)| = \bar \Lambda_n([\partial \phi]).
\end{equation}
Thus, Proposition\ \ref{prop:disk} implies that $\sum_j D_\phi(y_j)$
is strictly greater than $\sum_j |d_\phi(y_j)|$ provided that
$\Lambda_n([\partial \phi])$ is strictly greater than $\bar
\Lambda_n([\partial \phi])$.  For example, if $[\partial \phi] =
c_1c_2c_1^{-1}c_2^{-1}$, then $\phi$ takes values $y_1$ or $y_2$
at least twice, even though $y_1$ and $y_2$ are of degree zero.

For our applications we shall want to consider maps $\nuvec: D^2
\rightarrow S^2$ from the two-disk into the two-sphere.  Let $R =
\{\evec_0,\ldots, \evec_n\} \in \Rcal(\nuvec)$ denote a set of $n+1$
regular values of $\nuvec$.  In contrast to the case of maps to the
plane, the algebraic degrees $d(\evec_j)$ are not determined by
$\partial \nuvec$, since the image of $\nuvec$ itself is determined
only up to whole coverings of $S^2$.  We can remove
this ambiguity by specifying the degree at one of the regular values,
eg
$ d_\nuvec(\evec_0) = d_0$.
We may identify the fundamental group $\pi_1(S^2 - R, \qvec)$ with the
free group $F_n(c_1,\ldots,c_n)$ on $n$ generators.  As above, we
specify that the generator $c_j$ corresponds to a closed loop 
which encircles $\evec_j$ once with positive orientation but
contains no other $\evec_k$'s, which determines the $c_j$'s up to
conjugacy.  $c_0$, which corresponds to a loop about
$\evec_0$, may be expressed as a product of the generators
$c_1$ through $c_n$ and their inverses.  In what follows, we write $b \sim c$ to denote that $b$
and $c$ are conjugate. 
In analogy with Proposition~\ref{prop:disk}, we have the following:
\begin{proposition}\label{prop:sphere}
Given $\nuvec: D^2\rightarrow S^2$ smooth, $R =
\{\evec_0,\ldots,\evec_{n}\} \subset \Rcal(\nuvec)$, and 
$\pi_1(\Rr^2-R,p) \backsimeq F_n(c_1,\ldots,c_n)$ with generators $c_j$ as
above, such that $c_0 \in F_n(c_1,\ldots,c_n)$.  Suppose that $d_\nuvec(\evec_0) = d_0$.  Then
\begin{equation}
  \label{eq:sphere}
  \sum_{j=0}^n D_{\nuvec}(\evec_j) \ge |d_0| +
\min_{\substack{ g_1,\ldots,g_{r+|d_0|}\sim c_0,\\ 
    h_1,\ldots,h_r\sim c_0^{-1}}}
\Lambda_n([\partial \nuvec]g_1\cdots g_{r+|d_0|}h_1 \cdots h_r).
\end{equation}
\end{proposition}
While it is straightforward to compute the spelling length of a given
element $g$, evaluating (\ref{eq:sphere}) may not be
as straightforward.

\subsection{Reflection-symmetric homotopy classes in a
  prism}\label{sec:refl-symm-homot}
A crude way to obtain a lower bound for the Dirichlet
energy of tangent unit-vector fields on $P$ is to estimate the contributions
from nonoverlapping balls centred on each vertex. 
Let $\nvec$ denote a smooth tangent unit-vector field on $P$ with invariants
 $h$.  Let $\vvec^a$
denote the $a$th vertex of $P$, and let $O^a \subset S^2$ denote the
set of directions about $\vvec^a$ subtended by $P$.  
 For $r$ less than the length of any of the edges coincident at
 $\vvec^a$, define $\nuvec^a_r:O^a \rightarrow S^2$ by
\begin{equation}
  \label{eq:nua}
  \nuvec^a_r(\evec) = \nvec(\vvec^a + r\evec).
\end{equation}
Up to parameterisation, $\nuvec^a_r$ describes the restriction of
$\nvec$ to a spherical cleaved surface of radius $r$.
We have that
\begin{equation}
  \label{eq:nuvec_estimate}
  |\grad \nvec(\vvec^a + r\evec)|^2 \ge \frac{1}{r^2}
   |(\nuvec^a_r)^{\bf '}(\evec)|^2 \ge 2 
\frac{1}{r^2}
   |\det (\nuvec^a_r)^{\bf '}(\evec)|,
\end{equation}
where the last inequality follows from the same reasoning as in
(\ref{eq:local_ineq}).
Then
\begin{equation}
  \label{eq:crude_lb_1}
  E(\nvec) \ge \sum_a \int_0^{R^a} \Omegaabs^{a}_r \,dr,
\quad \text{where}\   \Omegaabs^{a}_r = \int_{O^a}  |\det
({\nuvec^a_r})^{\bf '}|\, 
d{\bf e}
\end{equation}
and the $R_a$'s are chosen so that $R_a + R_b \le |\vvec_a -
\vvec_b|$.

The quantity $ \Omegaabs^{a}_r$ is just the unsigned area of $\imag \nuvec^a_r$.    The
unsigned area of $\imag \nuvec^a_r \cap \Sigma^\sigma$ 
is at least the area of $\Sigma^\sigma$ times the minimal absolute
degree of the regular values in $\Sigma^\sigma$.  Thus we have that
\begin{equation}\label{eq:Omega_est_1}
  \Omegaabs^{a}_r \ge \sum_\sigma  \min_{\evec \in \Rcal(\nuvec^a_r) \cap
  \Sigma^\sigma} D_{\nuvec_r^a}(\evec) A^\sigma.
\end{equation}
Noting that $ D_{\nuvec_r^a}(\evec_0) \ge |w^{a\sigma_0}|$
for all $\evec_0 \in \Rcal(\nuvec^a_r) \cap
  \Sigma^{\sigma_0}$,
we may apply
Proposition\ \ref{prop:sphere} to (\ref{eq:Omega_est_1}) to obtain
\begin{equation}\label{eq:Omega_est_2}
  \Omegaabs^{a}_r \ge |w^{a\sigma_0}| A^{\sigma_0} +  \sum_{\sigma\ne
 \sigma_0}
 \min_{\substack{g_1,\ldots,g_{r+|w^{a\sigma_0}|}\sim c_{\sigma_0},\\ 
    h_1,\ldots,h_r\sim c_{\sigma_0}^{-1}}}
\Lambda_{s-1}([\partial \nuvec]g_1\cdots g_{r+|w^{a\sigma_0}|}h_1 \cdots h_r)\,
A^\sigma
\end{equation}
($s$ in (\ref{eq:Omega_est_2}) 
is the number of sectors).
We note that it follows from (\ref{eq:Omega_est_1}) and (\ref{eq:deg_and_Deg})  that
\begin{equation}\label{eq:Omega_est_3}
  \Omegaabs^{a}_r \ge 
\sum_{\sigma} |w^{a\sigma}|\,A^\sigma.
\end{equation}

For certain homotopy classes of tangent unit-vector fields on a rectangular
prism, $R$, one can show that the estimate (\ref{eq:Omega_est_3})
based on spelling lengths leads to an improvement of the
lower bound of  Theorem~\ref{thm:1}.  Let
\begin{equation}
  \label{eq:prism_desc}
  R = \{\rvec \, | \, 0 \le r_j \le L_j, j = x,y,z\},
\end{equation}
where for 
convenience we have chosen coordinates with the origin at one of the
vertices and axes parallel to the edges.
By convention, we take $L_x \ge L_y \ge L_z$.
In this case, the sectors are the coordinate octants of $S^2$ with area $A^\sigma = \pi/2$.

{\it Reflection-symmetric} homotopy
classes on $R$ are the homotopy classes of tangent unit-vector fields
which are invariant under reflections through the mid-planes of the
prism,
\begin{equation}
  \label{eq:reflection-symmmetric_n}
  \nvec(x,y,z) = \nvec(L_x-x,y,z) = \nvec(x,L_y-y,z) = \nvec(x,y,L_z-z).
\end{equation}
In this case, the wrapping numbers at two vertices ${a}$ and ${ \bar
  a}$ related by a single reflection differ by a sign;
\begin{equation}
  \label{eq:ref_sym_w}
  w^{a\sigma} = -w^{\bar a \sigma}.
\end{equation}
Thus, the wrapping numbers about the origin determine all the rest, and for
simplicity we denote these by $w^\sigma$.  The prism, and
reflection-symmetric configurations in particular, will also feature
in Sections~\ref{sec:upper-bound-prism} and
\ref{sec:exist-regul-minim}.

To estimate $E^\inff_R(h)$ for  reflection-symmetric $h$, it suffices to
consider tangent unit-vector fields $\nvec$ which are themselves
reflection symmetric. From (\ref{eq:crude_lb_1}) and
(\ref{eq:Omega_est_3}) it follows that
\begin{equation}
  \label{eq:rs_lb_1}
 E^\inff_R(h)  \ge 4\pi  \sum_{\sigma} |w^{\sigma}| L_z,
\end{equation}
which
coincides with the lower bound (\ref{eq:E_-}) of Theorem\
\ref{thm:1} (a minimal
connection in this case is obtained by pairing vertices at the endpoints of the
(shortest) $L_z$-edges of $R$).
However, by using the estimate (\ref{eq:Omega_est_2}) instead of (\ref{eq:Omega_est_3}), we get
the following.  
\begin{theorem}\label{thm:refl-symm-homot-1}
{\rm \cite{mrz2007b}}
  Let $h$ be a nonconformal reflection-symmetric homotopy class in
  $R$.  Let $\sigma^+$ denote the sector with largest positive
  wrapping number, denoted $W^+$, and let $\sigma^-$ denote the sector
  with largest (in magnitude) negative wrapping number, denoted $W^-$.
Let
  \begin{equation}
    \label{eq:Delta}
    \Delta(h) = \max\left(W^+ - \sum_{\sigma\in adj(\sigma^+) \, |\,
    w^\sigma > 0} w^\sigma - \chi,\ \
    |W^-| - \sum_{\sigma \in \adj(\sigma^-)\, | w^\sigma < 0}
    |w^\sigma| - \chi, \ \ 0\right),
  \end{equation}
where $\adj(\sigma)$ denotes the set of (three) octants
adjacent to (ie, sharing an edge with) $\sigma$, and $\chi$ is equal to 0
or 1 depending on the signs of the edge orientations and kink numbers.
Then
\begin{equation}
  \label{eq:rs_lb_2}
 E^\inff_R(h)  \ge 4\pi \left( \sum_{\sigma} |w^{\sigma}| + 2\Delta(h)\right)L_z.
\end{equation}
\end{theorem}
For typical nonconformal homotopy classes, $\Delta(h) > 0$.

\section{Upper bound in a prism}\label{sec:upper-bound-prism}  

In Theorem~\ref{thm:0}, one obtains an equality for the infimum
Dirichlet energy for a prescribed set of point defects,
rather than just a lower bound, by constructing a sequence
$\nvec^{(j)}$ whose energies approach $8\pi L(\Pcal(d),\Ncal(d))$.  It
can be shown that a subsequence $\nvec^{(k)}$ approaches a constant
away from lines joining the paired defects in a minimal connection
(here assumed unique), while $|\grad \nvec^{(k)}|^2$ approaches a
singular measure supported on these lines 
\cite{brecorlie}.  
For
tangent unit-vector fields on $P$, the boundary conditions preclude
such a construction; $\nvec$ is required to vary across the faces of
$P$, and therefore throughout its interior.  However, by constructing tangent
unit-vector fields which saturate the local inequality
(\ref{eq:local_ineq}) over most of $P$, we can produce upper bounds
for the Dirichlet energy with the same scaling with homotopy
invariants as the lower bound of Theorem~\ref{thm:1}.
Details are given in 
\cite{mrz2004b, majthesis, mrz2007, mrz2007b}.

Here in outline is a procedure for constructing such configurations.
Fix a set of values $h$ of the homotopy invariants.
As in Section~\ref{sec:refl-symm-homot}, let
$O^a\subset S^2$ denote the set of directions about the vertex
$\vvec^a$ subtended by $P$.  Define spherical cleaved surfaces
\begin{equation}\label{eq:C^a}
  C^a = \{ \vvec^a + r^a
\evec\, | \,\evec \in S^2 \},
\end{equation}
where $r^a$ is taken to be less than half the length of the smallest edge
coincident at $\vvec^a$ (so that the $C^a$'s do not intersect).  Specify
$\nvec$ on $C^a$ so as to satisfy tangent boundary conditions with
wrapping numbers given by $h$, and take $\nvec$ to be constant along 
rays from $C^a$ to $\vvec^a$.  It remains to define $\nvec$ on
$\Phat$, 
the (closed) domain obtained by excising the cones
between the $C^a$'s and $\vvec^a$'s.  The boundary of $\Phat$
is composed of i) the $C^a$'s and ii) the faces of $P$ truncated by
the $C^a$'s.  Extend $\nvec$ smoothly to these
truncated faces so as to satisfy tangent boundary conditions.  Choose
a point $\pvec$ in the interior of $\Phat$.  Along rays from $C^a$ to
$\pvec$, take $\nvec$ to be constant.  Along rays from each
truncated face to $\pvec$, rotate the values of $\nvec$ out of the
tangent plane to the outward normal.
There
emerges  a discontinuity  at $\pvec$, but this is easily removed.
If $\nvec$ is specified on the $C^a$'s to be 
conformal or anticonformal, except possibly on a small subset where its
derivative is suitably controlled, then the local inequality
(\ref{eq:local_ineq}) is saturated throughout most of $P$, and
the Dirichlet energy can be shown to be proportionate to the lower
bound of Theorem~\ref{thm:1}
independently of $h$.

The main difficulty in carrying out this procedure is in defining
$\nvec$ on the $C^a$'s.  
Let
$\nuvec^a:O^a \rightarrow S^2$ be given by $\nuvec^a(\evec) =
\nvec(\vvec^a + r^a \evec)$ (similarly to (\ref{eq:nua})).
We note that
$O^a$ is a geodesic polygon on $S^2$; its sides are arcs of the
great
circles 
of directions tangent to the faces of $P$ which
are coincident at
$\vvec^a$.  Tangent boundary conditions require that $\nuvec^a$
maps each side of $O^a$ into the great circle containing it.
If $h$ is conformal with respect to $\vvec^a$ (the anticonformal and
nonconformal cases are discussed below), we are led to the following:
\begin{problem}\label{prob:conformal_map}
  Find conformal maps on $S^2$ which preserve a given set of
  geodesics.
\end{problem}
Restricting the domain of such a  map to $O^a$ yields a
candidate for
$\nuvec^a$.

In the case of the rectangular prism $R$, Problem~\ref{prob:conformal_map} is
readily solved.  There are three geodesics which meet at right angles,
and which may be taken to be the great circles about $\xhat$, $\yhat$,
and $\zhat$.    Under the stereographic projection $\evec
\mapsto w = (e_x + ie_y)/(1+e_z)$ from $S^2$ to the extended complex
plane, these are mapped to the
real axis, imaginary axis and unit circle respectively. 
Problem~\ref{prob:conformal_map} becomes one of finding
locally analytic functions $f(w)$ such that i) $f(w)$ is real when $w$
is real, ii) $f(w)$ is imaginary when $w$ is imaginary, and iii)
$|f(w)| = 1$ when $|w| = 1$.  Property i) implies that $f$ is real; ii) then
implies that $f$ is odd; iii) then implies that $f(1/w) =
1/f(w)$.  Therefore, if $w_*$ is a zero of $f$, then $-w_*$ and
$\wbar_*$ are zeros, while $1/\wbar_*$ is a pole.  Restricting to $f$
to be meromorphic, we may conclude that $f$ is rational of the form
\begin{multline}
    \label{eq:w(z)}
    f(w) = \pm w^{2m+1}
  \prod_{j=1}^a\left(
  \frac{w^2 - r_j^2}{r_j^2 w^2 - 1}
  \right)^{\rho_j}
  \prod_{k=1}^b\left(
  \frac{w^2 + s_k^2}{s_k^2 w^2 + 1}
  \right)^{\sigma_k}\times\\
\times  \prod_{l=1}^c\left(
  \frac{(w^2 - t_l^2)(w^2 -  \tbar_l^2)}
  {(t^2_l w^2 - 1)({\tbar}_l^2w^2  - 1)}
  \right)^{\tau_l}.
\end{multline}
The $r_j$'s denote the real zeros ($\rho_j = 1$) and poles ($\rho_j =
-1)$ of $f$ between $0$ and $1$; the $s_k$'s, the imaginary zeros
and poles of $f$ (according to $\sigma_k = \pm 1$) between $0$ and $i$
; and the $t_l$'s, the complex zeros and poles of $f$ (according to
$\tau_l = \pm 1$) with modulus less than one and argument between 0 and
$\pi/2$.  

The parameters in (\ref{eq:w(z)}) can be chosen to realise any
admissible set of conformal (ie, nonpositive) wrapping numbers.
Anticonformal topologies can be realised by replacing $w$ with $\bar
w$.  Nonconformal topologies can be produced by modifying $f$
in a small neighbourhood to be anticonformal and smoothly
interpolating between the conformal and anticonformal domains.

Let
\begin{equation}
  \label{eq:E^-}
  E^-_P(h) = \sum_\sigma 2 A^\sigma L(\Pcal^\sigma(h), \Ncal^\sigma(h))
\end{equation}
denote the lower bound of Theorem~\ref{thm:1}.
\begin{theorem}\label{thm:2}
{\rm \cite{mrz2007}}
  Let $R$ denote a rectangular prism with sides of length $L_x \ge L_y
  \ge L_z$ and largest aspect ratio $\kappa = L_x/L_z$.  Then
  \begin{equation}
    E_R^\inff(h) \le C \kappa^3 
E^-_P(h)
  \end{equation}
  for some constant $C$ independent of $h$ and $L_x$, $L_y$, $L_z$.
\end{theorem}
In the proof of Theorem~\ref{thm:2}, the positions of the zeros and
poles of the conformal map (\ref{eq:w(z)}) must be chosen carefully to ensure
that the bound is achieved.

For reflection-symmetric conformal homotopy classes (cf (\ref{eq:reflection-symmmetric_n})), a simpler
construction leads to an improved result, in which $C =1$ and
$\kappa^3$ is replaced by $(L_x^2 + L_y^2 + L_z^2)^{1/2}/L_z$.
\begin{theorem}\label{thm:3}
{\rm \cite{mrz2004b}}
  Let $R$ denote a rectangular prism with sides of length $L_x \ge L_y
  \ge L_z$ and $h$ a reflection-symmetric homotopy class which is
  conformal about one of the vertices.  Then
  \begin{equation}
    E_R^\inff(h) \le \frac{(L_x^2 + L_y^2 + L_z^2)^{1/2}}{L_z} 
E^-_P(h).
  \end{equation}
\end{theorem}
Theorem\ \ref{thm:3} extends to nonconformal reflection-symmetric homotopy
classes, provided $E^-_P(h)$ is replaced by the lower bound given by
Theorem\ \ref{thm:refl-symm-homot-1} 
\cite{mrz2007b}.

\section{Existence and regularity of local minimisers: numerical
  results}\label{sec:exist-regul-minim} Using direct methods, one
might expect to establish the existence of a global minimiser of the
Dirichlet energy for tangent unit-vector fields on $P$.  The existence
of (continuous) local minimisers in a given homotopy class $h$,
however, is more difficult to address.  The homotopy classes are not
weakly closed, so that the existence of such local minimisers is not
guaranteed; $E_P^\inff(h)$ may not be realised (just as the infimum
energy for a prescribed set of point defects in $\Rr^3$ is not
realised).  We also recall the Hardt-Lin phenomenon 
\cite{hl4} --
global minimisers of the Dirichlet energy may have interior
singularities, even when continuous unit-vector fields are admissible.
If continuous local minimisers exist, then one would like to analyse
their regularity 
\cite{su82, su, duzaar2004, moser}.  

Questions about the
existence and regularity of continuous local minimisers of given
homotopy type appear to be open for the problems we are considering.
Below we describe some numerical results which suggest that, for some
homotopy classes, smooth minimisers always exist, while for others,
they may exist or not depending on the geometry of $P$.

\begin{figure}
\begin{center}
\includegraphics[totalheight=0.24\textheight,
viewport= 70 54 553 418,clip
] {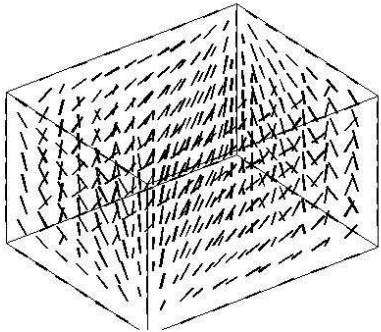}
\caption{Unwrapped configuration in class $h_0$}
\label{fig: unwrapped}
\end{center}
\end{figure}

The first examples concern two reflection-symmetric homotopy classes
in a rectangular prism, denoted here by $h_0$ and $h_1$, which are
both conformal with respect to one of the vertices (and, therefore,
conformal or anticonformal with respect to the others).  Details are
given in 
\cite{mrz2004b}.  
$h_0$ is the simplest possible, in which
there is a single nonzero wrapping number equal to $-1$, so that
$\nvec$ takes values in a single octant of $S^2$.  The restriction of
$\nvec$ to a spherical cleaved surface corresponds to the conformal
map given by $f_0(w) = w$ (cf (\ref{eq:w(z)})).  Such a configuration
is shown in Fig.\ \ref{fig: unwrapped}.
The lower bound for the infimum energy of Theorem\ \ref{thm:1} is $4\pi
L_z$.  The upper bound of Theorem~\ref{thm:3} 
can be improved in this case by explicit
evaluation of the Dirichlet energy for a trial configuration, yielding
\begin{align}\label{eq:unwrapped energy}
  E_R^\inff(h_0) &< 8  \frac {L_y L_z}{L_x^2}
  F_2\left(1,\half,\half,
{\textstyle \frac{1}{2}},
{\textstyle \frac{1}{2}}, -\frac{L_y^2}{L_x^2}, -\frac{L_z^2}{L_x^2}
\right)\nonumber \\
& \ \ + (x \rightarrow y \rightarrow z) + 
(x \rightarrow z \rightarrow y),
\end{align}
where $F_2(\alpha,\beta,\beta',\gamma,\gamma';s,t)$ is the Appell
hypergeometric function 
\cite{gradshteyn}.  
For a unit cube, we get
the bounds
\begin{equation}
   \label{eq:unit_cube}
   12.5 \lesssim E_R^\inff(h_0) \lesssim 15.3
\end{equation}
We computed minimisers numerically using two methods, namely solution of
the Euler-Lagrange equation (using FEMLAB, a commercial PDE solver)
and gradient descent.  The converged energies from both
methods agree, giving approximately $14.8$.  The converged unit-vector
field is indistinguishable from Fig.\ \ref{fig: unwrapped} at the resolution
shown, and appears to be regular away from the vertices.

\begin{figure}
\begin{center}
\includegraphics[totalheight=0.21\textheight,
viewport= 70 40 555 418,clip]{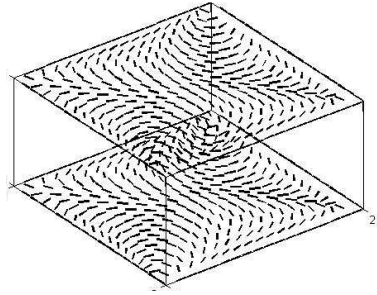}
\caption{A configuration $\nvec$ in $h_1$, generated by the conformal map
  $f(w) = w(w^2+s^2)/(s^2w^2 + 1)$ for $s = .5$.  $\nvec$ describes a
  3/4-turn about each vertex in the top and bottom face.}
\label{fig: wrapped}
\end{center}
\end{figure}

The homotopy class $h_1$ is the next simplest among the
reflection-symmetric conformal classes.  There are three nonzero
wrapping numbers equal to $-1$ in contiguous octants, so that $\nvec$
takes values in three-quarters of a hemisphere. 
The restriction of $\nvec$ to a spherical cleaved surface corresponds to the conformal map
\begin{equation}\label{eq:wrapped}
  f_1(w) = w\frac{w^2 + s^2}{s^2 w^2 + 1}.
\end{equation}
Such a configuration is shown in Fig.\ \ref{fig: wrapped}.
On the $xy$-faces of
the prism, $\nvec$ executes a three-quarter turn about each vertex,
corresponding to kink numbers of $\pm 1$; the kink numbers on the
other faces all vanish.
As the parameter $s$ approaches $1$ from below, half-turns becomes concentrated
along the $y$-edges, while $f_1$ approaches 
$f_0$ away from the $y$-edges. The corresponding family of
configurations $\nvec_s$ is weakly but not strongly continuous with
respect to $s$ (an example of the fact that $h_1$ is
not weakly closed).

\begin{figure}
\begin{center}
\centerline{\includegraphics[totalheight=0.19\textheight]{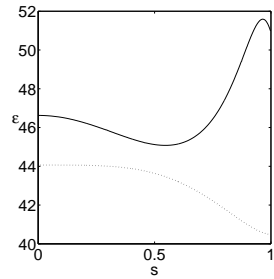}}
\caption{Scaled energy
 $\epsilon = E/(L_x L_y L_z)^{1/3}$ of the conformal configuration
  $w(w^2 + s^2)/(s^2 w^2 + 1)$. Solid curve: $L_x = 20$, $L_y = 10$,
  $L_z = 1$.  Dashed curve: $L_x = L_y = L_z = 1$.}
\label{fig: energy plot}
\end{center}
\end{figure}

Both numerical methods indicate that $h_1$ supports a smooth local
minimiser for sufficiently thin slabs ($L_x/L_y, L_x/L_z \lesssim 1/10$),
while for aspect ratios closer to unity, the numerical solution
converges to the  minimiser in $h_0$.  Some insight into this
behaviour is provided by computing the Dirichlet energy of trial
configurations characterised by the one-parameter
family~(\ref{eq:wrapped}), as shown in Fig.~\ref{fig: energy plot}.  For a cube (dashed curve),
the energy approaches a minimum as $s$ approaches 1, corresponding to
a configuration in which half-turns concentrate along the $y$-edges.  For $L_x = 20$,
$L_y = 10$, $L_z = 1$ (solid curve), the energy has a minimum for $s$
between $0$ and $1$, corresponding to a smooth configuration.
Note that concentration along the shortest ($L_z$)-edges
(which support the minimal connection) is not compatible with
the topology, as the nonzero kink numbers lie in the $xy$-faces.
Analogous arguments suggest that reflection-symmetric
homotopy classes with two or more nonzero kink numbers do not 
contain smooth minimisers (for these classes, concentration along the
shortest edge is compatible with the topology).  However, it is
conceivable that more non-reflection-symmetric homotopy
classes (for which minimal connections do not necessarily pair
vertices along edges) 
support smooth local minimisers.

\begin{figure}
\includegraphics[totalheight=0.24\textheight,
viewport= 80 60 426 318, clip]{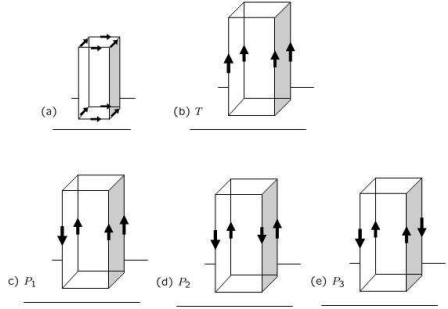}
\caption{Edge orientations for the four PABN configurations. (a) 
  Orientations on the horizontal edges are the same for all. (b) Tilted
  profile $T$.  $\nvec$ points up on all vertical edges.
(c) Planar profile $P_1$.  $\nvec$ points down on a single vertical
  edge  (d) Planar profile $P_2$.  $\nvec$ points down on a pair of
  adjacent vertical edges  (e) 
Planar profile  $P_3$. $\nvec$ points down on a pair of opposite
  vertical edges.}
\label{fig: pabn_figs}
\end{figure}

The last numerical example is an idealised model of the PABN device.
Details are given in 
\cite{mrz2007c}.
In fact, the model lies outside the class of problems we have
considered so far; the domain is not a polyhedron, and the boundary
conditions are not purely tangent.  We take the PABN to consist of a
rectangular post of square cross-section centred on the bottom surface
of a rectangular cell of square cross-section, as in Fig.\ 
\ref{fig: pabn_figs}.  In keeping with the device dimensions, the cell
height is taken to be three times the cell width, and the cell width
to be twice the post width.  The height of the post is variable.  Boundary
conditions are dictated by material characteristics of the substrates.  
Tangent boundary
conditions apply on the bottom substrate and on the post, while normal
boundary conditions are appropriate for the top substrate.  Periodic boundary
conditions are imposed on the vertical sides of the cell, simulating a
two-dimensional array of cells supporting
the same nematic configuration (at a given time) and comprising a
single pixel.

We consider four simple homotopy classes, in which the kink numbers
are zero and the trapped areas taken to have their minimal allowed
values.  The orientation of $\nvec$ on the horizontal edges of the
post are fixed, as in Fig\ \ref{fig: pabn_figs}(a). The classes are
distinguished by the relative orientations of $\nvec$ on the vertical
edges of the post.  Up to symmetry, there are four distinct
possibilities.  For the tilted class $T$, the orientation on all four
vertical edges is the same.  The other three classes, called planar,
are obtained by taking the orientation to be opposite on,
respectively, one of the vertical edges (the $P_1$ class), two
adjacent vertical edges ($P_2$), and two opposing vertical edges
($P_3$).  Configurations in $T$ exihibit a large vertical component
$n_z$ in the region around the post.  In configurations in $P_1$ --
$P_3$, $n_z$ is suppressed by the change in orientation between the
vertical edges.

\begin{figure}[h]
\includegraphics[totalheight=0.28\textheight
]{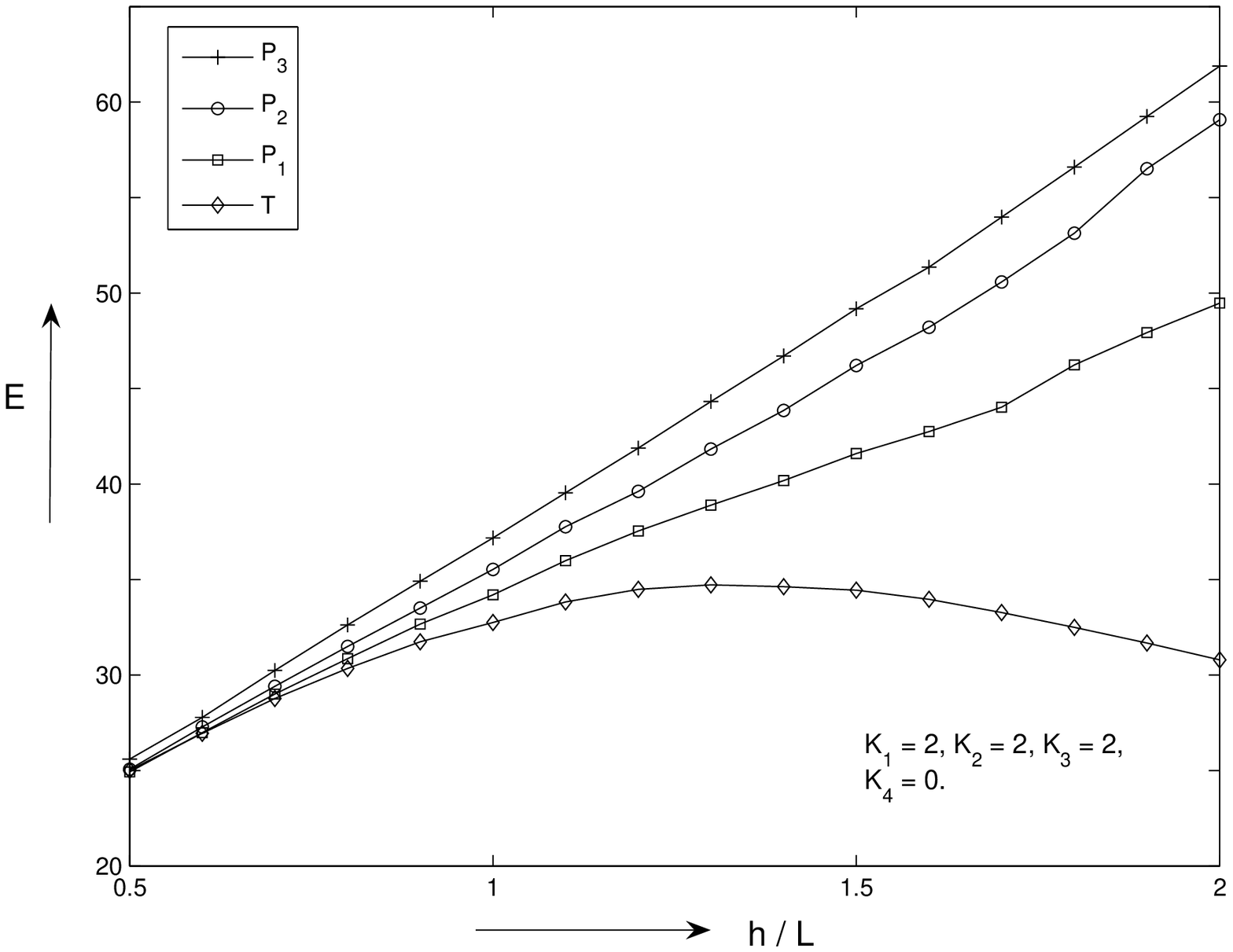}
\caption{Energies in the PABN cell as a function of post height}
\label{fig: pabn_plot}
\end{figure}

Local minimisers for each of these homotopy classes were computed
using FEMLAB for a range of post heights.  The converged
configurations appear to be smooth away from the vertices of the
post.  In Figure\ \ref{fig: pabn_plot} we plot the converged energies
of the local minimisers as a function of post height.  The tilted
class has the lowest energy, which is consistent with experimental observations
which show that the liquid crystal always relaxes into the high-tilt
state when cooled from the isotropic state 
\cite{kg2002}.
The computations support the hypothesis that the bistable states of
the PABN are topologically distinct.

\vspace{1cm}
\noindent{\it Acknowledgments.} We thank CJP Newton, our co-author on
\cite{mrz2007c}, 
for many helpful discussions and, along with A
Geisow, for stimulating our interest in these problems.
AM was partially supported by an EPSRC/Hewlett-Packard Industrial CASE
Studentship.  AM and MZ were partially supported by 
EPSRC grant EP/C519620/1.









\bibliographystyle{plain}
\bibliography{review4}


\end{document}